# Fictitious Supercontinent Cycles

J. Marvin Herndon
Transdyne Corporation
San Diego, CA 92131 USA
mherndon@san.rr.com

**Abstract:** Descriptions of phenomena, events, or processes made on the basis of problematic paradigms can be unreasonably complex (e.g. epicycles) or simply wrong (e.g. ultraviolet catastrophe). Supercontinent cycles, also called Wilson cycles, are, I submit, artificial constructs, like epicycles. Here I provide the basis for that assertion and describe published considerations from a fundamentally different, new, indivisible geoscience paradigm which obviate the necessity for assuming supercontinent cycles.

When individuals seek to describe phenomena, events, or processes within the framework of a problematic paradigm, the explanations proffered are generally more complex, if not physically impossible, than subsequent, corresponding explanations within a different, but more-correct paradigm. For example, in the Earth-centered Ptolemaic universe paradigm, the apparent motion of planets, in particular their retrograde motions, were described by complex epicycles (Figure 1). For another example, in the classical, pre-quantum physics paradigm, an ideal black body at thermal equilibrium was calculated to emit radiation with essentially infinite power in the shorter wavelengths, the so-called ultraviolet catastrophe, a physical impossibility. Within the now-known, more-correct paradigms, those and other phenomena can be explained logically, causally, and with greater simplicity, without invoking complex, *ad hoc* assumptions.

Geological literature contains a plethora of papers [1-3] dealing with various aspects of so-called "supercontinent cycles", also called "Wilson cycles", the idea that before Pangaea, there were a series of supercontinents that each formed and then broke apart and separated before colliding again, re-aggregating, and suturing into a new supercontinent in a continuing sequence. Here, I suggest that "supercontinent cycles" are artificial constructs, like epicycles, attempts to describe geological phenomena within the framework of problematic paradigms.

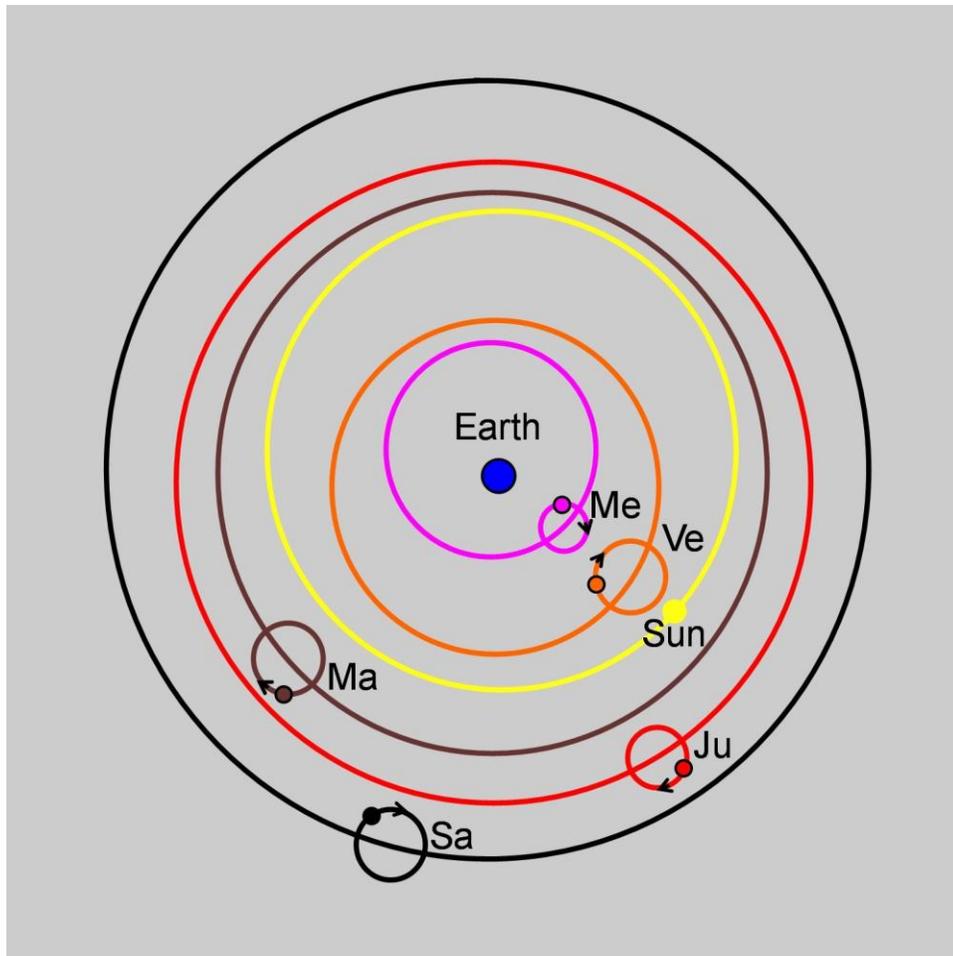

**Figure 1.** Epicycles were able to explain apparent retrograde motion of planets in the problematic Earth-centered Ptolemaic universe paradigm.

Concepts of planetary formation generally fall into one of two categories that involve either (1) condensation at high-pressures, hundreds to thousands times the pressure of our atmosphere at Earth's surface; or (2) condensation at very, very low-pressures.

Since the 1960s, the scientific community almost unanimously concurred that Earth formed from primordial matter that condensed at a very low-pressure, ca. $10^{-4}$ atm. [4, 5]. The 'planetesimal hypothesis' was 'accepted' as the 'standard model of solar system formation'. The idea was that dust would condense from the gas at this very low pressure. Dust grains would collide with other grains, sticking together to become progressively larger grains, then pebbles, then rocks, then planetesimals and finally planets [6, 7]. However, as I discovered, there is an inherent flaw in that paradigm [8-10].

The inner planets all have massive cores, as known from their high relative densities. I was able to show by thermodynamic calculations that the condensate of primordial matter at those very low pressures would be oxidized, like the Orgueil C1/CI meteorite wherein virtually all elements are combined with oxygen. In such low-pressure, low-temperature condensate, there would be essentially no iron metal for the massive-cores of the inner planets, a contradiction to the observation of massive-core planets.

The planetesimal hypothesis, *i.e.*, the 'standard model of solar system formation', is not only problematic from the standpoint of planetary bulk-density, but necessitates additional *ad hoc* hypotheses. One such necessary hypothesis is that of a radial solar-system temperature gradient during planetary formation, an assumed warm inner region delineated by a hypothetical 'frost line' between Mars and Jupiter; ice/gas condensation is assumed to occur only beyond that frost line. Another such necessary hypothesis is that of whole-planet melting, *i.e.*, the 'magma ocean', to account for core formation from essentially undifferentiated material.

Beginning in the 1960s, the plate tectonics hypothesis was developed and 'accepted' by many as the paradigm to explain Earth dynamics. The topography and magnetic striations of the seafloor are explained well by basalt being extruded at mid-oceanic ridges, moving across the ocean expanse, and disappearing into trenches. Plate tectonics, in a manner consistent with the planetesimal hypothesis, explains the mechanism and fate of "subducted" ocean floor basalt slabs as being part of mantle convection cells that act as great conveyer belts recycling ocean floor basalt into the mantle.

In 1931, Holmes [11] introduced the concept of mantle convection (Figure 2) as a motive force for Wegener's continental drift [12]. In Holmes' mantle convection idea, the rocky part of the Earth is assumed to circulate in great loops, like endless conveyer belts, dragging the continents along. The assumption of mantle convection is a critical component of plate tectonics, not only for seafloor spreading, but also for continental movement; continent masses are assumed to ride atop assumed convection cells, much as Holmes envisioned for continental drift. In plate tectonics, plate collisions are thought to be the sole mechanism for fold-mountain formation. Indeed, the occurrence of mountain chains characterized by folding which significantly predate the breakup of Pangaea is the primary basis for assuming the existence of supercontinent cycles with their respective periods of ancient mountain-forming plate collisions.

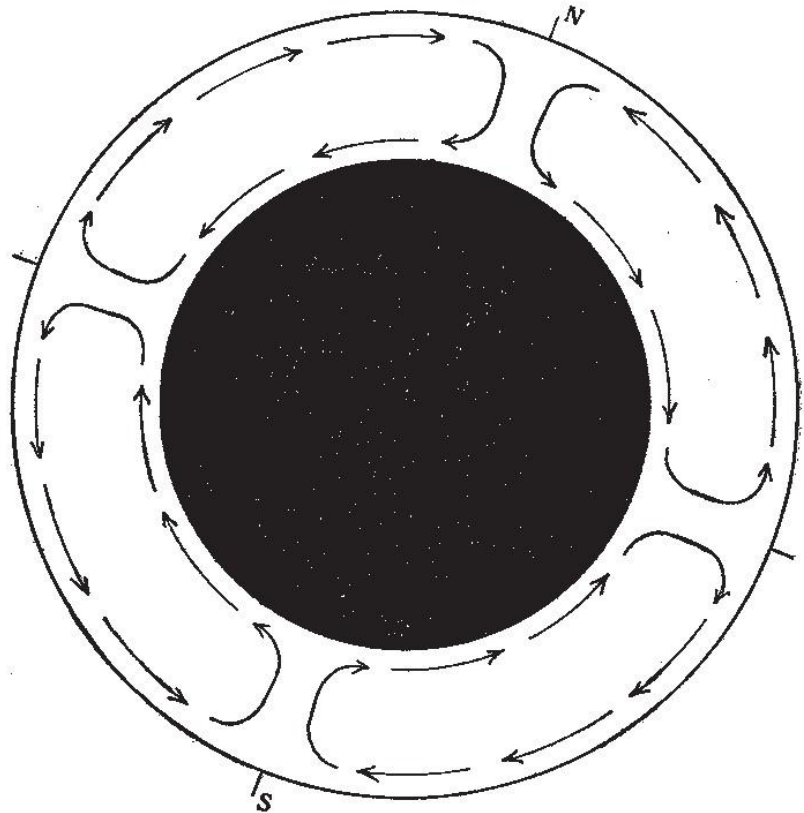

**Figure 2.** Schematic representation of mantle convection by Holmes [11]. Reproduced with permission of the Geological Society of Glasgow.

The assumption of mantle convection critically underlies virtually all aspects of plate tectonics including supercontinent cycles, but as I disclosed [13, 14], there is a serious problem. The Earth's mantle is 62% denser at the bottom than at the top [15] (Figure 3). The small amount of thermal expansion at the bottom (<1%) cannot overcome the 62% higher density at the mantle's bottom. Sometimes attempts are made to obviate the 'bottom heavy' prohibition by adopting the tacit assumption that the mantle behaves as an ideal gas with no viscous losses, *i.e.*, 'adiabatic'. But the mantle is a solid that does not behave as an ideal gas as evidenced by earthquakes occurring at depths as great as 660 km.

In the absence of mantle convection, plate tectonics is without a valid scientific basis. But that should not be surprising as there are other problems. For example, nowhere in the literature of plate tectonics is presented a logical, causally related explanation for the fact that about 41% of Earth's surface is continental rock (sial) with the balance being ocean floor basalt (sima). Absent mantle convection, there is no motive force for driving supercontinent cycles. The reasonable conclusion one must draw is, as in the case of epicycles, there must exist a new and fundamentally different geoscience paradigm which obviates the problems inherent in plate tectonics and in planetesimal Earth formation and yet is capable of better explaining observed geological features.

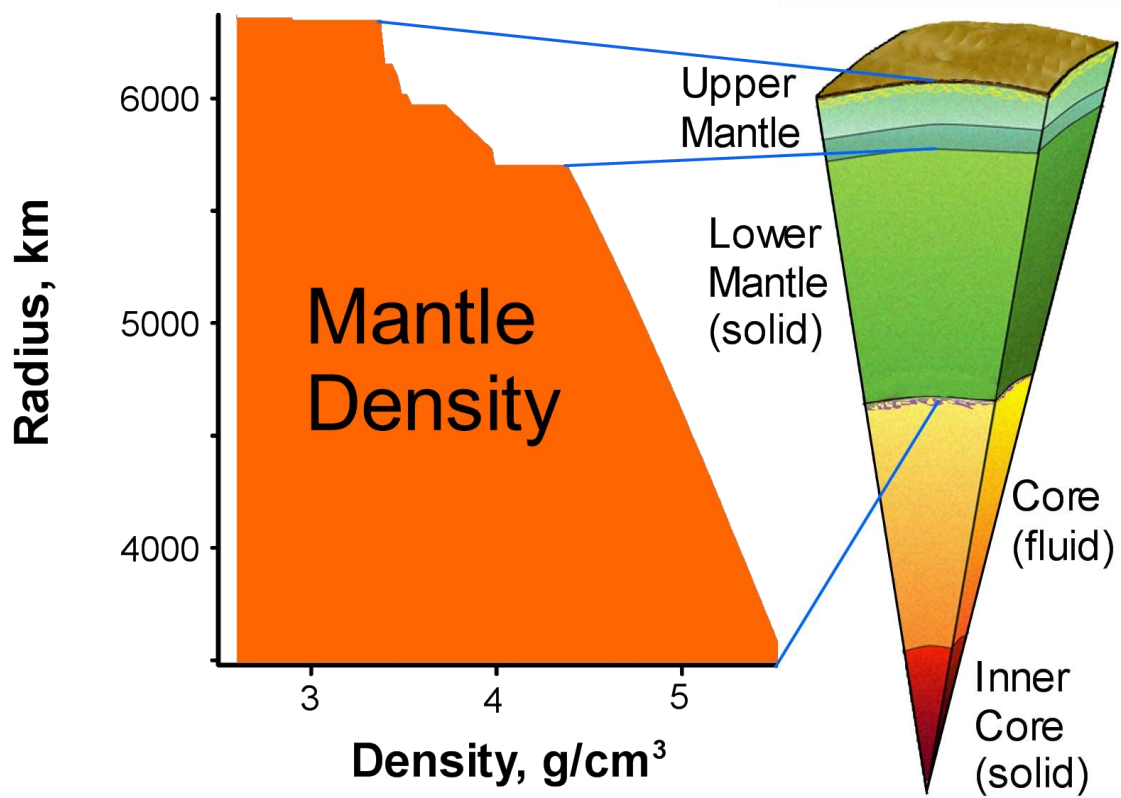

**Figure 3.** Density as a function of radius in the Earth's mantle [15].

I have disclosed a new indivisible geoscience paradigm, called Whole-Earth Decompression Dynamics (WEDD), that begins with and is the consequence of our planet's early formation as a Jupiter-like gas giant and which permits deduction of: (1) Earth's internal composition and highly-reduced oxidation state; (2) Core formation without whole-planet melting; (3) Powerful new internal energy sources, protoplanetary energy of compression and 'georeactor' nuclear fission energy; (4) Mechanism for heat emplacement at the base of the crust; (5) Georeactor geomagnetic field generation; (6) Decompression-driven geodynamics that accounts for the myriad of observations attributed to plate tectonics without requiring physically-impossible mantle convection, and; (7) A mechanism for fold-mountain formation that does not necessarily require plate collision. The latter obviates the necessity to assume supercontinent cycles.

I have described the details and implications of Whole-Earth Decompression Dynamics in a number of scientific articles [9, 10, 14, 16-21] and books [22-26]. Briefly, as first suggested by Eucken [27], Earth's core rained-out by condensing from solar matter at high-pressures and high-temperatures, followed by the more-volatile silicates. Complete condensation, I submit, led to Earth's early formation as a Jupiter-like gas giant. The weight of 300 Earth-masses of gas bearing down on the rocky kernel of Earth compressed the rocky portion to about 64% of its present diameter, sufficient compression for a solid continental-rock layer to cover the entire rocky part of the planet. After removal of the gases by solar T-Tauri eruptions, the enormous gravitational energy of compression, stored during the Jupiter-like phase, became available to power later decompression and its resulting geodynamic activity; what remained was a solid Earth, smaller than at present, whose rocky surface consisted entirely of continental rock (sial), without ocean basins. Eventually, internal pressure became sufficiently great to begin to crack the 100% closed contiguous shell of continental-rock that I call *Ottland*, in honor of Ott Christoph Hilgenberg, who first conceived of its existence [28].

According to Whole-Earth Decompression Dynamics, the geology of planet Earth is primarily the consequence of two processes: (1) The progressive formation of surface cracks to increase surface area in response to decompression-increased planetary volume, and; (2) The progressive adjustment of surface curvature in response to decompression-increased planetary volume.

Regarding (1) above, driven by the stored energy of protoplanetary compression, augmented by georeactor nuclear fission energy, surface cracks are of two types: *primary* with underlying heat sources, and *secondary* that lack heat sources. Basalt extruded from primary cracks migrates and eventually falls into and in-fills secondary cracks, a process that develops ocean basins and yields understanding of seafloor magnetic striations and topography even better than plate tectonics and without requiring mantle convection.

Regarding (2) above, illustrated by the demonstration in Figure 4, curvature changes of the continental-rock surface, necessitated by decompression-increased planetary volume, as I have disclosed [21], can be accommodated by buckling, breaking, and falling over upon itself. Formed in this manner, mountain ranges characterized by folding contain their 'extra' surface area within present continental boundaries and do not necessarily require or imply continent collision. Thus the primary basis for assuming supercontinent cycles is obviated.

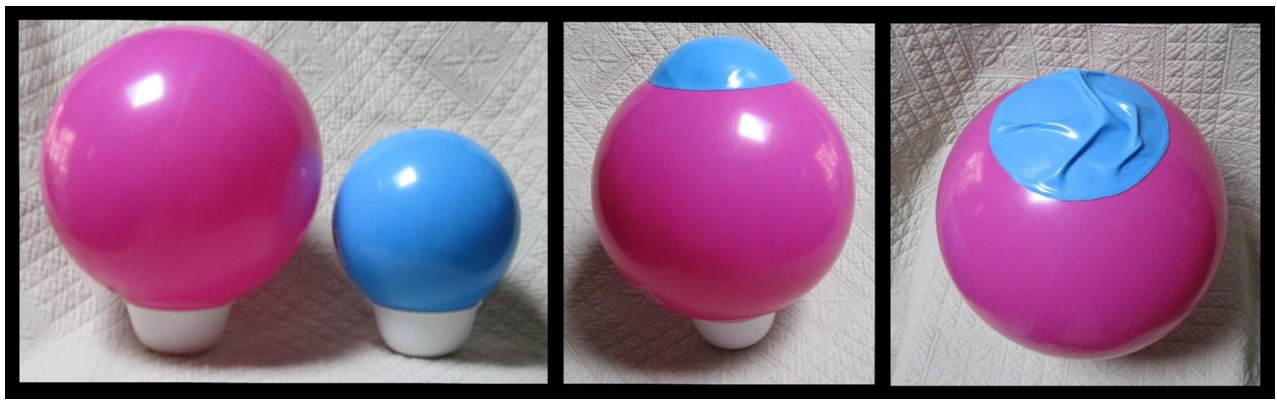

**Figure 4.** Demonstration illustrating the formation of fold-mountains as a consequence of Earth's early formation as a Jupiter-like gas giant [21]. On the left, two balls representing the relative proportions of 'present' Earth (pink), and 'ancient' Earth (blue) before decompression. In the center, a spherical section, representing a continent, cut from 'ancient' Earth and placed on the 'present' Earth, showing: (1) the curvature of the 'ancient continent' does not match the curvature of the 'present' Earth and (2) the 'ancient continent' has 'extra' surface area confined within its fixed perimeter. On the right, tucks remove 'extra' surface area and illustrate the process of fold-mountain formation that is necessary for the 'ancient' continent to conform to the curvature of the 'present' Earth. Unlike the ball-material, rock is brittle so tucks in the Earth's crust would break and fall over upon themselves producing fold-mountains.

Models of supercontinents engaged in hypothetical Wilson cycles typically make use of problematic paleomagnetic calculations. As I have shown [20], whole-Earth decompression can lead to significant errors in magnetic paleo-latitude calculations. Moreover, paleo-pole calculations, used to imply continent rotations, are without meaning due to changes in Earth-radius. No means of supercontinent locomotion, fold-mountain formation without the necessity of collisions, significant errors in magnetic paleo-latitude calculations, and the invalidity of magnetic paleo-pole calculations all together call into question the entire concept of supercontinent cycles. Fictitious supercontinent names, such as Rodinia, Columbia, and even Pangaea, will eventually pass into history along with planetary epicycles. The challenge for geologists will be to discover the true sequence of fragmentation beginning with Ottland and continuing to the present and to discover the nature of Earth's surface throughout that progression.

**Dedication:** This work is dedicated to the memory of Lynn Margulis (1938-2011) who repeatedly urged and insisted that it should be written.